\documentclass[aps,prc,groupedaddress,twocolumn,showpacs]{revtex4-1}
\pdfoutput=1
\usepackage{graphicx}
\usepackage[breaklinks, pdftex]{hyperref}
\usepackage[breaklinks]{hyperref}
\usepackage[english]{babel}
\usepackage{booktabs}
\graphicspath{{../figures/}}
\usepackage{epstopdf}
\usepackage{rotating}
\usepackage{multirow}
\usepackage{amsmath}
\usepackage{dcolumn}

\begin{document}

\title{A reanalysis of radioisotope measurements of the $^9$Be$(\gamma,n)^8$Be cross section}

\author{Alan E. Robinson}
\affiliation{Fermi National Accelerator Laboratory, Batavia, Illinois, USA}
\email{fbfree@fnal.gov}

\date{June 17, 2016}

\begin{abstract}
The $^9$Be$(\gamma,n)^8$Be reaction is enhanced by a near threshold $1/2^+$ state.  Contradictions between existing measurements of this reaction cross section affect calculations of astrophysical r-process yields, dark matter detector calibrations, and the theory of the nuclear structure of $^9$Be.  Select well-documented radioisotope $^9$Be$(\gamma,n)$ source yield measurements have been reanalyzed, providing a set of high-accuracy independently measured cross sections without the large systematic errors from recent beamline experiments \cite{BeOxs, Utsunomiya2}.  A single-level Breit-Wigner fit of these corrected measurements yields are $E_R=1736.8(18)$~keV, $\Gamma_\gamma=0.742(25)$~eV, and $\Gamma_n=252(17)$~keV for the $1/2^+$ state, excluding a virtual state solution.
\end{abstract}

\pacs{26.30.Hj, 25.20.-x, 27.20.+n}

\maketitle

The near threshold $1/2^+$ state of the $^9$Be$(\gamma,n)^8$Be reaction is important for several processes in nuclear and astrophysics.  This cross section is used to calculate the formation rate of $^9$Be via the $^4$He$(\alpha n,\gamma)^9$Be reaction, one of the most important light-element reactions for r-process nucleosynthesis \cite{Be9_importance}.  Neutrons from $^{88}$Y/Be and other radioisotope sources using the $^9$Be$(\gamma,n)^8$Be reaction near threshold are being used widely for dark matter detector energy scale and yield calibrations \cite{CDMS_Be, PICO-60} as these low energy neutrons closely mimic dark matter recoils in detectors \cite{Juan_YBe}.

Recent experimental \cite{Utsunomiya2} and theoretical studies of $^9$Be suggest that three-cluster dynamics are required to describe its photodissociation.  This could imply that the astrophysical $^9$Be production proceeds via a single step, with a much larger production rate at low energies than that calculated by the two-step process via $^8$Be \cite{Be9_cluster1, Be9_cluster2}.  Three-cluster models calculations using the complex scaling method do not find a resonant $1/2^+$ state \cite{Be9_virtual, three-body2}, but a virtual $1/2^+$ state would not be found by these calculations.  A virtual state has a complex energy eigenvalue of \cite{*[{For a detailed review of nuclear reaction resonances, see }][] R-matrix},
\begin{gather}
\label{eq:virtual}
E_\lambda = (E_R - S_n) + i\Gamma/2 \\
|E_\lambda| \begin{cases} < 0 & \text{ virtual state}\\ > 0 & \text{ resonance state}\end{cases} \notag
\end{gather}
defined by the real quantities $E_R$, the resonance energy, $S_n$, the neutron separation energy, and $\Gamma$, the resonance width.  These parameters defining the virtual or resonant nature of the $1/2^+$ state can be measured from the position and shape of the near threshold peak of the $^9$Be$(\gamma,n)$ cross section.

\begin{figure}
\centering
\includegraphics{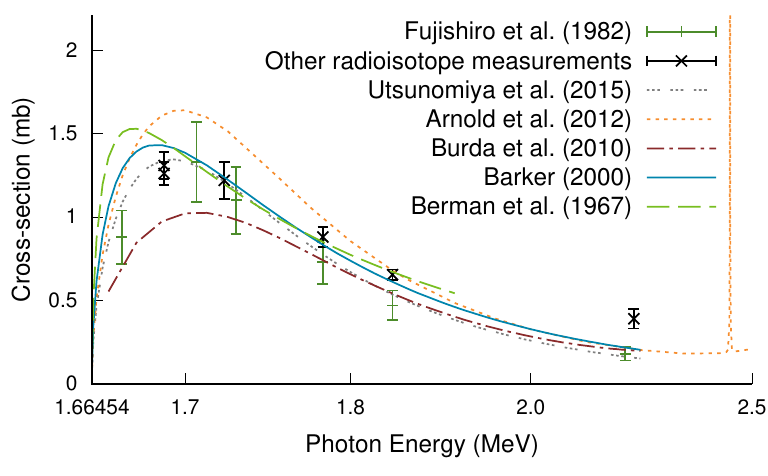}
\caption{\label{fig:th_ca_BeOxs}A selection of existing near-threshold cross section measurements of the $^9$Be$(\gamma,n)^8$Be reaction\cite{th_ca_ORNL, th_ca_ANL2, th_ca_Gibbons, th_ca_John_Prosser, th_ca_Fuji1982, BeOxs, Utsunomiya2, th_ca_Burda, th_ca_Berman}.  Lines show the cross section from fitted Breit-Wigner parameters. The parameters from \citet{th_ca_Barker} were fit to data from \cite{th_ca_Kuechler}.}
\end{figure}

There is little agreement between many measurements of the parameters of this $1/2^+$ state.  The yields of radioisotope photoneutron sources provide the simplest method to measure the cross section, a technique that has been used many times since 1935~\cite{th_ca_Chad1935, th_ca_Halban1938, th_ca_ANL, th_ca_ORNL, th_ca_ANL2, th_ca_Gibbons, th_ca_John_Prosser, th_ca_Fuji1982}.  These measurements use a limited set of precisely known photon energies and can use homogeneous and isotropic neutron detectors with well understood sensitivity to the low-energy neutrons produced by the reaction.  Other measurements using bremsstrahlung photon beams~\cite{th_ca_Jakobsen, th_ca_Berman}, and more recently using inverse Compton photon beams~\cite{th_ca_Utsunomiya, BeOxs, Utsunomiya2}, such as the High-Intensity $\gamma$ Source (HI$\gamma$S), provide cross sections over a range of energies inaccessible to photoneutron sources while sacrificing energy resolution and simplicity in experimental design.  All these techniques rely on a comparison of absolute quantities.  Accurate cross section measurements require knowledge of the absolute photon source strengths, the neutron detection efficiencies, and the photon energy spectra.  The energy dependence and systematic uncertainty of the neutron detection efficiencies are improved when simple, homogeneous and isotropic neutron detectors are used.  The cross section can also be found using the inelastic scattering of charged particles~\cite{th_ca_Tucker, th_ca_Spencer, th_ca_Nguyen, th_ca_Clerc, th_ca_Kuechler, th_ca_Dixit, th_ca_Burda}.  These measurements require significant background subtraction and extrapolation to low momentum transfer to recover a cross section.  A selection of measured and evaluated cross sections in Figure~\ref{fig:th_ca_BeOxs} show disagreements of up to 60\%, with the strongest disagreements among recent measurements.

\begin{table*}
\centering
\begin{ruledtabular}
\begin{tabular}{l c c c c}
& \multicolumn{2}{c}{from John and Prosser} & \multicolumn{2}{c}{Recalculated} \\
\cline{2-3}\cline{4-5}\\[-1.7ex]
     & Correction & Uncertainty & Correction & Uncertainty \\
Item & (\%) & (\%) & (\%) & (\%) \\
\hline\\[-1.7ex]
1. Gamma-ray source strength & & & & \\
(a) peak-to-total ratio                 & ---    & 5.0    & $+3.88$ & 3.5 \\
(b) subtraction of other $\gamma$-ray lines & ---    & 2.0    & $+0.42$ & 2.0 \\
(c) half-life of $^{124}$Sb             & ---    & 0.8    & $-0.8$ & 0.1 \\[2pt]
2. Neutron source strength &&&& \\
(a) absolute strength of NBS source     & ---    & 3.0    & $+0.4$ & 0.85 \\
(b) neutron escape from MnSO$_4$ bath   & $-0.9$ & 0.3    & $-0.84$ & 0.05 \\
(c) capture of fast neutrons in bath    & $-2.8$ & 0.5    & $-2.34$ & 0.12 \\
(d) neutron counting statistics         & ---    & 0.8    & --- & 0.8 \\[2pt]
3. Finite source size                   & $-0.6$ & $<0.1$ & --- & --- \\[2pt]
4. Attenuation of gamma rays in Be shell & $+2.5$ & $<0.1$ & --- & --- \\[2pt]
5. Neutrons produced by other $\gamma$-ray lines
\footnote{Not applied when calculating neutron yields (see text).}
& $-4.4$ & 1.0 & $-2.49$ & --- \\[2pt]
6. Source geometry effects              &        &        & $+2.65$ & 2.0 \\
\cline{2-3}\cline{4-5}\\[-1.7ex]
Total Correction                        & $-6.2\%$ && 1.007 & \\
Overall uncertainty (square root of sum of squares) && $6.4\%$ && $4.7\%$ \\
\end{tabular}
\end{ruledtabular}
\caption{\label{tab:th_ca_John_Prosser}Corrections to the measured $^9$Be$(\gamma,n)$ cross section from an $^{124}$Sb/Be source, reproduced from Table I in \citet{th_ca_John_Prosser} and extended with recalculated corrections.  These corrections reconstruct the measured neutron production cross section compared to that of a 1.69~MeV photon through a thin beryllium target.  The finite source size (3) and gamma attenuation (4) effects are combined and reported with other source geometry effects (6) in an MCNPX-Polimi simulation in the reevaluation.  See text for descriptions of other recalculated corrections.}
\end{table*}

Radioisotope measurements of the $^9$Be$(\gamma,n)^8$Be cross section using simple detector designs should have low systematic uncertainties.  However, discrepancies among some of these measurements are as severe as those from more complicated experiments at accelerator facilities.  Most of the radioisotope measurements were performed in the 1940s, 50s, and 60s, before high precision neutron standards, cross sections, and simulation programs were available.  Fortunately, several of these experiments have been well documented with results that are traceable to modern calibrations.  This paper will apply corrections to these originally measured cross section values to construct a trusted set of high-accuracy measured radioisotope source neutron yields near the $^9$Be$(\gamma,n)$ threshold.

\section{Radioisotope Measurements}

\subsection{John and Prosser}

\citet{th_ca_John_Prosser} used the MnSO$_4$ bath technique \cite{th_ca_NBS1} to measure the yield of a $^{124}$Sb/Be source and traced their measured neutron yield to NBS-1, the world's most precisely calibrated neutron standard.  The strength of their $^{124}$Sb source was measured using a scintillating NaI crystal.

A list of corrections to the measurement was provided in \citet[][Table I]{th_ca_John_Prosser}, and is reproduced with corrections in Table~\ref{tab:th_ca_John_Prosser}.  The largest correction and uncertainty is to the peak to total ratio of the NaI crystal.  \citeauthor{th_ca_John_Prosser} cite an unpublished calibration of a a 2" thick 1.75" diameter crystal using radioisotopes with simple decay schemes.  Their  uncertainty, and an uncertainty common to NaI crystal measurements at the time, was due to their efficiency calibration and its extrapolation in photon energy \cite{Nucleonics, IDO}.  Today, photon propagation Monte Carlo methods can accurately predict the peak to Compton ratio of a detector with known dimensions.  An MCNP simulation of the detector with a 1.41 MeV threshold gives a ratio of 0.2244 versus the originally used ratio of 0.216.  A 3.5\% uncertainty (5\%$/\sqrt{2}$) in the photon source strength is retained to account for uncertainties subdominant to those specified by \citeauthor{th_ca_John_Prosser}.

In 1962, only the 1.69~MeV and the 2.09~MeV lines were well established in the high energy $^{124}$Sb photon spectrum \cite{Sb_decay}.  \citeauthor{th_ca_John_Prosser} subtracted from the measured photon yield the contribution from the Compton tail from the 2.09-MeV line.  Several percent and sub-percent intensity sub-threshold photon lines have been found since 1962 that lie within the NaI detector resolution of the 1.69-MeV peak  \cite{th_ca_sb124}.  Assuming a NaI(Tl) detector with 5.4\% energy resolution at the peak, an additional 0.42\% background subtraction is applied.  The $^{124}$Sb half-life has been revised to 60.20(3)~days from 60.4$\pm$0.2~days.  As the neutron counting significantly preceded the photon counting, the decay correction was increased.

To calibrate the neutron yield against NBS-1, an intermediate $^{226}$Ra/Be $(\alpha,n)$ source was used.  Correction factors for this comparison were recalculated using MCNPX-Polimi simulations, replacing the original analytic corrections made by \citeauthor{th_ca_John_Prosser}.  Photoneutrons were generated in the simulation by generating photons and simulating the $^9$Be$(\gamma,n)^8$Be reaction.  An MCNP library \cite{Alan_PRC} based on the cross section measured by \citet{BeOxs} was used.  The $^{226}$Ra/Be neutron spectrum used to generate source particles in the simulation was calculated using the JENDL-AN/05 evaluated $^9$Be$(\alpha,n)$ cross section \cite{JENDL_AN} and a modified version of the SOURCES-4C program \cite{SOURCES, Alan_thesis}.    Neutron propagation was modelled using the ENDF/B-VII.1 cross section libraries~\cite{ENDF-VII}.  The ratio of the number of neutrons captured on manganese and the number of neutrons produced was calculated from these simulations and used to compare the $^{226}$Ra/Be and $^{124}$Sb/Be neutron yield measurements.  The largest difference in yield is due to the increased probability of neutron capture on sulfur and oxygen for high energy neutrons from the $^{226}$Ra/Be source.

The absolute neutron yield of the NBS-1 neutron source has been revised slightly upward by 0.4\% and the yield uncertainty reduced to 0.85\% since its 1955 calibration~\cite{th_ca_NBS1, th_ca_NBS3}.  

\citeauthor{th_ca_John_Prosser}, and most other neutron yield measurements, present cross sections after correcting for the neutron production from sub-dominant photon energies using other known points of the $^9$Be$(\gamma,n)$ cross section, such as with Item 5 in Table~\ref{tab:th_ca_John_Prosser}.  Radioisotope source neutron yields, the number of neutrons produced per decay, may be alternatively expressed as the sum of the photon branching ratios $\alpha_i$ and cross sections $\sigma_i$.  This reanalysis will present results in terms of radioisotope neutron yields.  For \citeauthor{th_ca_John_Prosser}'s $^{124}$Sb/Be measurement, a yield of $\sum_i \alpha_i \sigma_i=0.678(32)$~mb is found.  This expression ignores the contribution of bremsstrahlung photons from high energy $\beta$ decays, or neutron production from high energy betas that reach the beryllium.  The bremsstrahlung and beta contribution to the neutron yield was calculated \cite{inel_e} and found to be negligible for $^{124}$Sb and all other isotopes considered by \citeauthor{th_ca_John_Prosser}, \citeauthor{th_ca_Gibbons}, and \citeauthor{th_ca_ORNL}.  This and all other reanalyzed yields and yield ratios are given in Table~\ref{tab:result}.

\citeauthor{th_ca_John_Prosser} compared their $^{124}$Sb/Be source to $^{28}$Al/Be and $^{206}$Bi/Be sources.  As their neutron emission rate was too weak to measure using the MnSO$_4$ bath method, a `Long Counter' was used for the comparison.  MCNPX-Polimi simulations of the Harwell IV Long Counter~\cite{th_ca_AERE} found that the counter had equal sensitivity (within 1\%) to neutrons from each of the three sources.  \citeauthor{th_ca_John_Prosser}'s 3\% correction to the sensitivity of neutrons from $^{28}$Al/Be was reduced to 0.7\%.  No other corrections to the cross section beyond those in the original paper were added.  The reported cross sections were converted into the neutron yield ratios shown in Table~\ref{tab:result}.

\subsection{Gibbons \textit{et al.}}

\citet{th_ca_Gibbons} measured both $^{^{124}}$Sb/Be and $^{88}$Y/Be neutron sources to high precision using different techniques from those of \citeauthor{th_ca_John_Prosser}.  The decay rate of their sources were determined using a $4\pi$ ionization counter.  Their neutron source strengths were measured using a 5-foot diameter graphite moderating sphere and BF$_3$ thermal neutron detectors.  The neutron count rate was calibrated against a source traceable to NBS-1.

\citet[][Table I]{th_ca_Gibbons} had calculated several corrections to the their source yield.  These corrections were recalculated, and found to be almost entirely in agreement with the original analysis.  Only the source strength of the NBS neutron source was changed, increasing the yield by 0.4\% and reducing its uncertainty to 0.85\%.

\begin{table}
\centering
\begin{ruledtabular}
\begin{tabular}{l c c c}
Measurement & Isotope & Yield  & Ratio to \\
            &         & or Yield Ratio & original \\
\hline\\[-1.5ex]
\citeauthor{th_ca_John_Prosser} & $^{124}$Sb & $0.678(32)$~mb & 1.079 \\
                                & $^{28}$Al/$^{124}$Sb & $1.311(43)$ & 0.978 \\
                                & $^{206}$Bi/$^{124}$Sb & $0.620(30)$ & 1 \\[4pt]
\citeauthor{th_ca_Gibbons} & $^{124}$Sb & $0.669(29)$~mb & 1.004 \\
                           & $^{88}$Y & $0.660(29)$~mb & 1.004 \\[4pt]
\citeauthor{th_ca_ORNL} & $^{24}$Na & $0.620(19)$~mb & 0.920 \\
                        & $^{72}$Ga & $0.171(5)$~mb & 0.792 \\[4pt]
\citeauthor{th_ca_Fuji1982} & $^{58}$Co & $0.0083(6)$~mb & 1.83 \\
                            & $^{105}$Ru & $0.00256(17)$~mb & 1.79 \\
                            & $^{65}$Ni & $0.0074(5)$~mb & 1.68 \\
                            & $^{28}$Al & $1.21(9)$~mb & 1.66 \\
                            & $^{88}$Y & $0.80(8)$~mb & 1.71 \\
                            & $^{38}$Cl & $<0.135(14)$~mb & $<1.69$ \\
\end{tabular}
\end{ruledtabular}
\caption{\label{tab:result} Reanalyzed source yields, $\sum_i \alpha_i \sigma_i$, for the given radioisotopes paired with $^9$Be.  Except for the $^{24}$Na/Be yield, these are used to fit the parameters of the $1/2^+$ state.  The values derived from \citet{th_ca_Fuji1982} shown above are freely scaled in the fit.  The yield for $^{38}$Cl includes $\leq10\%$ contributions from inelastic scattering of $\beta$-rays on beryllium and bremsstrahlung photons; an upper limit is set for the yield from $^{38}$Cl $\gamma$-rays.}
\end{table}

\subsection{Snell \textit{et al.}}

\citet{th_ca_ORNL} measured the neutron yields from beryllium and deuterium targets using $^{72}$Ga and $^{24}$Na radioisotopes.  Both isotopes produce photons above the 2.2~MeV deuterium dissociation threshold.  The absolute neutron yields and the ratio of beryllium to deuterium neutron yields were measured for each radioisotope.  The measured ratios can be compared to the well known modern cross section for the photodissociation of deuterium \cite{ENDF-VII}.  Both the reported absolute and the relative yields were reanalysed and found to be consistent within their final 5\% and 3\% uncertainties \cite{Alan_thesis}.  The yield measurement relative to the deuterium target is used in this reanalysis.

\citeauthor{th_ca_ORNL} measured neutrons by sampling epithermal neutrons in large volume of paraffin moderator using indium foil sandwiched between two cadmium foils.  The activated indium was counted using a thin-walled Geiger counter.  The neutron energy dependence of the neutron yield measurements of \citeauthor{th_ca_ORNL} has been validated using MCNPX-Polimi~\cite{MCNPX}.  One additional correction, for the thermalization of neutrons reentering the deuterium or beryllium from the surrounding moderator, leads to an small additional loss of efficiency.  The ratio of the detection efficiency for each source was recalculated.  The original uncertainty in the relative source activity is retained, and it dominates the total uncertainty of 3\%.

\subsection{Fujishiro \textit{et al.}}

The measurements of \citet{th_ca_Fuji1982} are the most recently published radioisotope measurements of the $^9$Be$(\gamma,n)^8$Be and have heavily influenced modern evaluations of the cross section~\cite{th_ca_Fuji_fit, th_ca_Burda, th_ca_NACRE}.  \citeauthor{th_ca_Fuji1982} used a reactor to irradiate a large variety of short-lived radioisotopes that produce photons with energies near the $1/2^+$ state, shown in Table~\ref{tab:th_ca_Fujisrc}.

\begin{table*}
\centering
\begin{ruledtabular}
\begin{tabular}{c c c c c c c c}
Isotope & $n$ count rate & Intensity & $n$ Det. Eff. & Branching Fraction & \multicolumn{2}{c}{Half-life} & Half-life \\
        & (s$^{-1}$) & (MBq) & (\%) & Correction & & & Correction \\[-2.7ex]
\cline{6-7}\\[-1.7ex]
& & & & & Fujishiro et al. & Revised & \\
        \hline\\[-1.5ex]
$^{58}$Co & 2.38(2) & 34.2(32) & 6.87 & $0.9945(1)\div0.9944$ & 70.8 d & 70.86(6) d & 0.995(5) \\
$^{105}$Ru & 0.403(10) & 18.8(16) & 6.79 & $0.473(5)\div0.481$ & 4.44 h & 4.44(2) h & 1.00(2) \\
$^{65}$Ni & 1.150(13) & 19.2(17) & 6.77 & $23.59(14)\div23.58$ & 2.52 h & 2.51719(26) h & 1.0058(5) \\
$^{28}$Al & 62.0(4) & 6.2(6) & 6.69 & 1 & 2.24 min & 2.245(2) min & 0.991(4) \\
$^{88}$Y & 7.01(4) & 1.12(13) & 6.55 & $0.937(3)\div0.9136$ & 106.6 d & 106.627(21) d & 0.9994(5) \\
$^{38}$Cl & 7.81(4) & 8.1(10) & 6.09 & $33.3(7)\div31.03$ & 37.14 min & 37.230(14) min & 0.989(2) \\
\end{tabular}
\end{ruledtabular}
\caption{\label{tab:th_ca_Fujisrc} Radioisotope sources of $> 1.67$~MeV photons used in \citet{th_ca_Fuji1982}.  The neutron count rate, neutron detection efficiency ($\varepsilon_n$), photon source intensity, and source intensity corrections are shown.  Half-lives and branching fractions are from relevant Nuclear Data Sheets \cite{th_ca_co58, th_ca_ru105, th_ca_ni65, th_ca_al28, th_ca_y88, th_ca_cl38} and from recent experiments \cite{th_ca_ru105_2, th_ca_cl38_2}.}
\end{table*}

In contrast to the other radioisotope measurements noted above, \citeauthor{th_ca_Fuji1982} used a non-homogeneous neutron detector: a ring of four BF$_3$ detectors embedded in a paraffin cylinder.  The neutron energy dependence of this detector was originally calculated using a one-dimensional Monte Carlo code and normalized to the flux of a $^{24}$Na--D$_2$O neutron source.  The relative neutron detection efficiency was recalculated using MCNPX-Polimi simulations of the three-dimensional geometry given by \citeauthor{th_ca_Fuji1982}.  The sensitivity, but not the model number, of the BF$_3$ counters used were given.  The new simulations use LND model 2025 BF$_3$ detectors filled to 500~torr for the 5.0~cps/nv thermal neutron detectors \cite{LND}.  The energy dependence of the neutron efficiencies calculated by MCNPX-Polimi match that found by \citeauthor{th_ca_Fuji1982}, but the new calculation finds an absolute efficiency that is much lower.  The calculated sensitivity to $^{24}$Na/deuterium neutrons is 6.2\% versus the $9.2\pm0.7\%$ measured and used as a calibration by \citeauthor{th_ca_Fuji1982}.

The cross section calculation of \citeauthor{th_ca_Fuji1982} was reproduced using the original neutron detection efficiency, background subtracted neutron count rates, source intensities, beryllium geometry, and calculation method presented in their paper.  Assuming a beryllium density of 1.85~g/cm$^3$, the recalculated cross sections are 12\% larger than those presented by \citeauthor{th_ca_Fuji1982}.  Given the normalization discrepancies in this calculation and the neutron detection efficiency, only the ratios of the neutron yields measured by \citeauthor{th_ca_Fuji1982} will be used.

The neutron yields are calculated using the MCNP calculated detector efficiency, and the neutron count rate, source intensities, and uncertainties given by \citeauthor{th_ca_Fuji1982}.  \citeauthor{th_ca_Fuji1982} included a 6\% uncertainty in the absolute normalization of their source intensities that is excluded when calculating relative source intensities in this reanalysis.  As the source intensities had been measured after they had decayed to $\sim 10^5$~Bq, a correction is made to account for updated half-life values.  The source intensities are also corrected to account for updated values of the branching fraction of the photon energy used to measure the source intensities.  These corrections are shown in Table~\ref{tab:th_ca_Fujisrc}.

Additional contributions to the neutron yield from high-energy $\beta$-rays and their bremsstrahlung photons were calculated.  The contributions are negligible ($<0.1\%$ of the total yield) for all measured radioisotopes except $^{38}$Cl.  The majority of $^{38}$Cl decays produce a $\beta$ with a 4.9~MeV endpoint energy.  Depending on the geometry of the source and encapsulation used by \citeauthor{th_ca_Fuji1982}, $\sim10\%$ of the measured neutron flux from their $^{38}$Cl/Be source could be caused by these high energy electrons.  A similarly unaccounted high-energy $\beta$ contribution had been incorrectly claimed by \citet{Fuji2} as evidence for the three-body breakup reaction of $^9$Be below the two-body threshold of $S_n=1664.54$~MeV \cite{PhysRevC.70.064611}.  No correction is applied for these additional neutron production mechanisms.  Instead, the measured $^{38}$Cl neutron yield is used an upper limit for the yield produced by the $^{38}$Cl $\gamma$ emissions.

\subsection{Other Radioisotope Measurements}

Other radioisotope measurements of the $^9$Be$(\gamma,n)^8$Be cross section exist that are not used in this reanalysis.  These measurements were either not calibrated against neutron standards, or insufficient information about the experiments were available to correct their measured yields.

\citet{th_ca_ANL2} measured the neutron yield from a $^{144}$Pr/Be source with its dominant emission at 2.185 MeV.  They used an unspecified NaI detector to measure the photon yield, and compared the strength of the 1.49-MeV $\gamma$ from $^{144}$Pr against the 1.38-MeV $\gamma$ from $^{24}$Na without correcting for the difference in detector efficiency between the two energies.  The change in detection efficiency depends by $\mathcal{O}(10\%)$ on the size and geometry of crystal used.  Their measurement is consistent with the $^{72}$Ga measurement of \citeauthor{th_ca_ORNL} and $^{38}$Cl measurement of \citeauthor{th_ca_Fuji1982} within this large uncertainty.

\citet{th_ca_ANL} measured the neutron yield of a set of six radioisotopes paired with beryllium and a subset of four radioisotopes paired with deuterium.  Their measurement of the neutron flux used a 'long counter' calibrated using a $^{226}$Ra-Be$(\alpha,n)$ neutron source.  This calibration source produces neutrons with far greater energies than the radioisotope sources and is used with a detector design whose sensitivity varies with neutron energy.  Their photon source activities were calculated from thermal neutron capture cross sections and the measured neutron flux of the nuclear reactors in which they were activated.  While improvements in the analysis of this measurement are possible, very large systematic uncertainties in both the neutron and photon yields would remain, and little information would be gained from a reanalysis.

Earlier measurement of $^9$Be$(\gamma,n)^8$Be source yields \cite{th_ca_Chad1935, th_ca_Halban1938,* [{Table III of }][{ and references therein.}] Wattenberg_1949} lack detailed information or the precision required for reanalysis.

\section{Fit}
\label{sec:fit}

\begin{figure}
\centering
\includegraphics{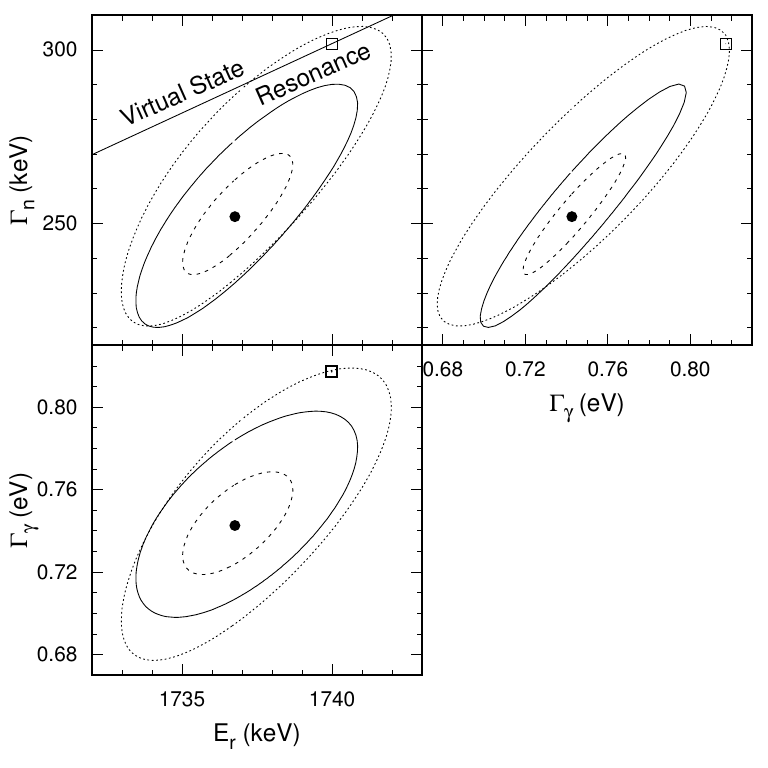}
\caption{\label{fig:fit_corr}Fit Breit-Wigner parameters to the $1/2^+$ state.  Best fit parameters (solid circles) are $E_R=1736.8(18)$~keV, $\Gamma_\gamma=0.742(25)$~eV, and $\Gamma_n=252(17)$~keV with a $\chi^2/$d.o.f. of $4.70/7$.  68\% (dashed) and 95\% (solid) confidence regions are shown.  A fit excluding the data from \citet{th_ca_Fuji1982} was also performed, with the 95\% confidence region shown by dotted lines.  A virtual state would exist if $\Gamma \approx \Gamma_n > E_R - S_n$ (see Equation \ref{eq:virtual}).  The best fit virtual state solution for the fit including all reanalyzed radioisotoope data is shown (open squares) and is excluded with 99.3\% confidence (2.7$\sigma$).  If the data from \citeauthor{th_ca_Fuji1982} is excluded, a virtual state is disfavored with 93\% confidence (1.8$\sigma$).  A script to calculate these likelihoods is available in the supplemental materials.} 
\end{figure}

\begin{figure}
\centering
\includegraphics{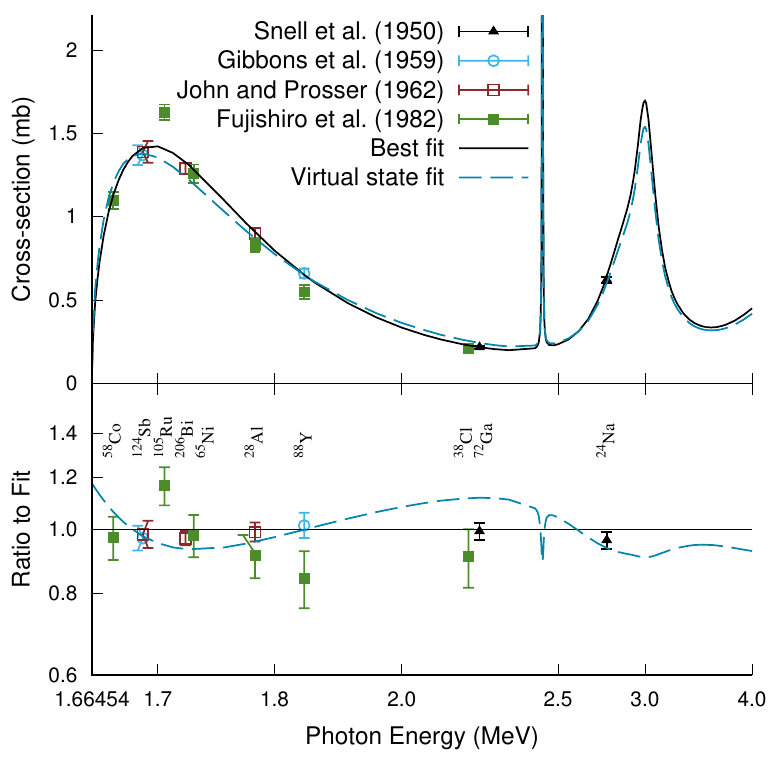}
\caption{\label{fig:th_ca_BeOxs2}Reanalyzed cross sections from radioisotope measurements of the $^9$Be$(\gamma,n)^8$Be cross section.  Only the cross sections for the highest intensity photon energy of each radioisotope are shown, assuming that ratios to the cross sections at other photon energies equal the best fit ratios.  The absolute yield of the measurements by \citet{th_ca_Fuji1982} are floated.  The 2.7~MeV measurement by \citet{th_ca_ORNL} of the $^{24}$Na/Be source yield was not included in the fit, but it is used to validate the contribution of higher energy resonances as measured by \citet{BeOxs}.}
\end{figure}

\begin{table}
\centering
\begin{ruledtabular}
\begin{tabular}{c c d d d}
$J^\pi$ & $E_R$ & \multicolumn{1}{c}{$\Gamma_\gamma$} & \multicolumn{1}{c}{$\Gamma_n$} & \multicolumn{1}{c}{$\beta_j$ to $^9$Be(g.s.)} \\
        & (keV) &    \multicolumn{1}{c}{(eV)}         & \multicolumn{1}{c}{(keV)}      & \multicolumn{1}{c}{(\%)} \\
\hline\\[-1.5ex]
$5/2^-$ & 2431 & 0.098 & 0.77 & 6 \\
$1/2^+$ & 2880 & 1.8 & 393 & 100 \\
$5/2^+$ & 3008 & 0.45 & 168 & 70 \\
$3/2^+$ & 4704 & 7.8 & 2419 & 38 \\
$3/2^-$ & 5590 & 15.7 & 1477 & 38 \\
\end{tabular}
\end{ruledtabular}
\caption{\label{tab:upper_res} Resonance parameters characterizing the contribution of high-energy resonances to the $^9$Be$(\gamma,n)$ cross section measured by \citet{BeOxs}.  These resonances are subtracted from the radioisotope data in order to fit the parameters of the near threshold $1/2^+$ state.  Note that the table values are used as a parameterization, and may not be physical.  In particular, the relatively narrow $3/2^-$ and $3/2^+$ resonances \cite{NDS_9} are being used to fit out the cluster dipole resonance discussed in Section~\ref{sec:comp}.  The widths of $3/2^+$ and $3/2^-$ resonances are multiplied by 1.57 from the values given by \citeauthor{BeOxs} in order to match their experimental data points.  A 10\% scale uncertainty on the summed cross section from these resonances is applied when fitting the parameters of the $1/2^+$ state.  The ground state branching ratios were selected to approximately match the observed ratio of thermal neutrons detected in two detector rings used by \citeauthor{BeOxs}. Uncertainties in the ground state branching ratios are not provided and may be large.
}
\end{table}

\begin{table*}
\centering
\begin{ruledtabular}
\begin{tabular}{c c D{.}{.}{2} D{.}{.}{5} d D{.}{.}{11} d}
Isotope & $t_{1/2}$ & \multicolumn{1}{c}{$E_\gamma$} & \multicolumn{1}{c}{$\alpha_i$} & \multicolumn{1}{c}{$E_n^{\text{c.m.}}$ to $^8$Be g.s.} & \multicolumn{1}{c}{Neutron yield} & \multicolumn{1}{c}{$^8$Be g.s. branching} \\
 & & \multicolumn{1}{c}{(keV)} & \multicolumn{1}{c}{(\%)} & \multicolumn{1}{c}{(keV)} & \multicolumn{1}{c}{($\sum_i \alpha_i \sigma_i$) (mb)} & \multicolumn{1}{c}{(\%)} \\
\hline \\[-1.5ex]
$^{58}$Co & 70.86(6) d & 1674.73 & 0.517(10) & 9.05 & 0.00584(24) & 100.0 \\[2pt]
$^{124}$Sb & 60.20(3) d & 1690.97 & 47.57(18) & 23.47 & 0.672(18) & 100.0 \\
                   && 2090.93 & 5.49(3) & 378.68 & 0.0145(6) & 99.5 \\
                   && \multicolumn{1}{l}{Others} & & & 0.00254(4) & 99.6 \\
                   && \multicolumn{1}{l}{Total} & & & 0.687(18) & 100.0 \\[2pt]
$^{105}$Ru & 4.44(2) h & 1698.17 & 0.0766(9) & 29.86 & 0.00109(3) & 100.0 \\
           &           & 1721.15 & 0.0299(3) & 50.27 & 3.93(14)\times10^{-4} & 100.0 \\
           && \multicolumn{1}{l}{Others} & & & 1.6(5)\times10^{-5} & 100.0 \\
           && \multicolumn{1}{l}{Total} & & & 0.00156(5) & 100.0 \\[2pt]
$^{206}$Bi & 6.243(3) d & 1718.7 & 31.9(5) & 48.10 & 0.424(16) & 100.0 \\
           &           & 1878.65 & 2.01(4) & 190.15 & 0.01062(26) & 99.9 \\
           && \multicolumn{1}{l}{Others} & & & 0.00580(22) & 99.4 \\
           && \multicolumn{1}{l}{Total} & & & 0.441(16) & 100.0 \\[2pt]
$^{65}$Ni & 2.51719(26) h & 1724.92 & 0.399(12) & 53.62 & 0.00513(23) & 100.0 \\[2pt]
$^{28}$Al & 2.245(2) min & 1778.99 & 100 & 101.64 & 0.909(19) & 100.0 \\[2pt]
$^{88}$Y & 106.627(21) d & 1836.06 & 99.2(3) & 152.33 & 0.646(7) & 100.0 \\
                   && 2734.0 & 0.71(7) & 949.8 & 0.0042(5) & 95.3 \\
                   && 3219.7 & 0.0070(20) & 1381.1 & 3.6(11)\times10^{-5} & 81.5 \\
                   && \multicolumn{1}{l}{Total} & & & 0.651(7) & 99.9 \\[2pt]
$^{38}$Cl & 37.230(14) min & 2167.40 & 44.4(9) & 446.58 & 0.101(5) & 99.1 \\
           && \multicolumn{1}{l}{Others} & & & 1.38(27)\times10^{-4} & 63.2 \\
           && \multicolumn{1}{l}{Total} & & & 0.102(5) & 99.0 \\[2pt]
$^{72}$Ga & 14.10(1) h & 1862.00 & 5.410(18) & 175.36 & 0.0309(4) & 99.9 \\
           &           & 2201.59 & 26.87(12) & 476.95 & 0.0585(25) & 98.9 \\
           &           & 2491.03 & 7.73(3) & 734.00 & 0.0179(8) & 94.8 \\
           &           & 2507.72 & 13.33(6) & 748.82 & 0.0318(14) & 95.3 \\
          && \multicolumn{1}{l}{Others} & & & 0.033(4) & 97.7 \\
           && \multicolumn{1}{l}{Total} & & & 0.172(6) & 97.8 \\[2pt]
$^{207}$Bi & 31.55(4) y & 1770.23 & 6.87(3) & 93.86 & 0.0660(16) & 100.0 \\[2pt]
$^{226}$Ra & 1600(7) y & 1729.60 & 2.878(8) & 57.78 & 0.0360(12) & 100.0 \\
                   && 1764.49 & 15.30(3) & 88.77 & 0.153(4) & 100.0 \\
                   && 1847.43 & 2.025(9) & 162.42 & 0.01244(14) & 100.0 \\
                   && 2204.06 & 4.924(18) & 479.14 & 0.0107(5) & 98.9 \\
                   && \multicolumn{1}{l}{Others} & 9.69 & & 0.0183(4) & 94.4 \\
                   && \multicolumn{1}{l}{Total} & & & 0.230(5) & 99.5 \\
\end{tabular}
\end{ruledtabular}
\caption{\label{tab:yields} Total neutron yields for reanalyzed and commercially available radioisotopes used in near threshold $^9$Be photoneutron sources.  For each major photon energy, the branching ratio to the $^8$Be ground state assuming the resonance parameters given in Table \ref{tab:upper_res}, and the energy of the outgoing neutron for that branch in the center of mass frame are given.  The neutron yield is the product of the $^9$Be$(\gamma,n)$ cross section $\sigma_i$ and the photon branching fraction $\alpha_i$ for each photon energy $E_\gamma$.  Photon yields, energies and half-lives, including those of decay chain daughter isotopes, are from~\cite{th_ca_co58, th_ca_y88, th_ca_sb124, th_ca_bi207, th_ca_ra226, th_ca_bi214}.  Neutrons with other final states, such as $\alpha + \alpha + n$ or $^8$Be(3030~keV)$+n$ will be produced at lower average energy and greater energy spread than those for the ground state transition.  Neutron yield uncertainties are propagated from the fit to the $1/2^+$ state and from photon branching ratios.
}
\end{table*}

The $^9$Be$(\gamma,n)^8$Be cross section was fit using least squares to a sum of isolated Breit-Wigner states.  The $1/2^+$ state was fit to the form
\begin{equation}
\sigma_{1/2^+}(E_\gamma)= \frac{\pi (\hbar c)^2}{4 E_\gamma^2}\frac{\Gamma_n \sqrt{\frac{E_\gamma-S_n}{E_R-S_n}} \quad \Gamma_\gamma \left(\frac{E_\gamma}{E_R}\right)^3}{(E_\gamma-E_R)^2+\frac{E_\gamma-S_n}{E_R-S_n}(\Gamma_n/2)^2}
\end{equation}
where the neutron separation energy $S_n=1664.54$~keV, $\Gamma_n$ and $\Gamma_\gamma$ are the neutron and photon channel widths, $E_R$ is the resonance energy, and $E_\gamma$ is the incoming photon energy.

The contribution to the cross section from other higher energy resonances may affect the measured width of the $1/2^+$ resonance.  \citet{BeOxs} provides the best direct measurement of the shape of the $^9$Be$(\gamma,n)$ cross section between 2~MeV and 4~MeV.  The contribution from higher energy resonances was fixed using their resonance parameterization shown in Table \ref{tab:upper_res}.  The widths of the fitted $3/2^+$ and $3/2^-$ resonances given by \citeauthor{BeOxs} were multiplied by 1.57 in order to match \emph{their reported experimental data} (see Table \ref{tab:upper_res} caption). Ten percent level discrepancies exist in the neutron detector efficiency calibration of \citeauthor{BeOxs} \cite{Arnold_eff}, and between their measurement and radioisotope measurements at low energy shown in Figure~\ref{fig:th_ca_BeOxs_comp}.  A smaller $5\%\pm3\%$ discrepancy exists between the $^{24}$Na/Be source yield measured by \citet{th_ca_ORNL} and the adopted contribution from high energy resonances.  Given these discrepancies, a 10\% scale uncertainty, larger than the 4.6\% systematic uncertainty given by \citeauthor{BeOxs}, was applied to the summed contribution from the higher energy resonances.

The photon energies, branching ratios, and reanalyzed measured neutron yield for each radioisotope measurement, shown in Table~\ref{tab:result}, were input to the fit.  The yields measured by \citet{th_ca_Fuji1982} were allowed to float freely.  The fit marginalized over the 10\%  uncertainty in the contribution from the higher energy resonances, and over the 0.85\% uncertainty in the NBS-1 neutron source strength.  The $\chi^2$ contribution of the $^{206}$Bi/Be and $^{28}$Al/Be yield measurements of \citet{th_ca_John_Prosser} were calculated using their ratio to the $^{124}$Sb/Be source yield.  The absolute source yields were used for all other measurements.  The $^{24}$Na/Be source yield from \citet{th_ca_ORNL} was excluded as its energy is well above the $1/2^+$ state.  Results from the fit are shown in Figures~\ref{fig:fit_corr}~and~\ref{fig:th_ca_BeOxs2}.  At the best fit point, the measurements of \citeauthor{th_ca_Fuji1982} and the higher resonance contribution are scaled down by $31.7\%$, and $1.1\%$ respectively.

A virtual state is disfavored by the fit with 99.5\% confidence.  As the measurements of \citeauthor{th_ca_Fuji1982} have large systematic uncertainties, require rescaling to match other radioisotope measurements, and have residuals in Figure \ref{fig:th_ca_BeOxs2} that hint at potential bias, a second fit was performed excluding their data.  This second fit still disfavors a virtual state interpretation, but only with 95\% confidence.

Recent measurements \cite{Utsunomiya2} and theoretical studies \cite{three-body2} suggest interference between positive parity states and a cluster dipole resonance near $E_\gamma=8$~MeV.  The effect of the interference should be small given the wide separation of these resonance energies, but even a small effect may reduce confidence in the exclusion of a virtual $1/2^+$ state.  Interference effects are neglected in this analysis.

For use in dark matter detector calibrations, the yields and uncertainties of individual commercially available photoneutron sources are given in Table~\ref{tab:yields}.

\section{\label{sec:comp} Comparison to Photon Beam Experiments}

\begin{figure}
\centering
\includegraphics{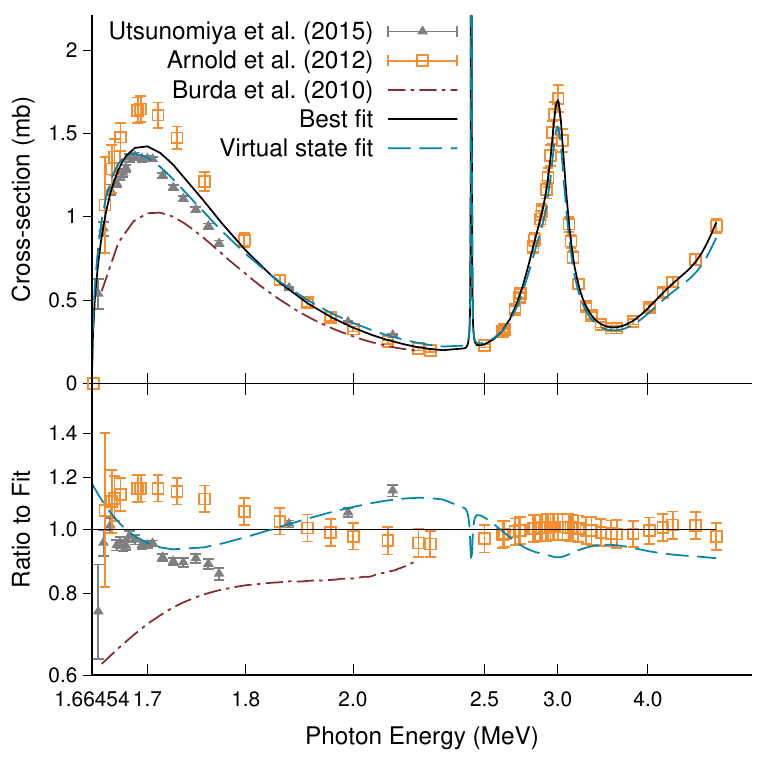}
\caption{\label{fig:th_ca_BeOxs_comp}Comparison of the fitted $^9$Be$(\gamma,n)^8$Be cross section to recent inverse Compton scattered photon beam \cite{BeOxs, Utsunomiya2} and electron scattering \cite{th_ca_Burda} measurements (see text).}
\end{figure}

Figure~\ref{fig:th_ca_BeOxs_comp} and Table~\ref{tab:peak} compare recent beamline measurements of the $^9$Be$(\gamma,n)$ reaction to the present reanalysis of radioisotope measurements.  Except for the measurements of \citeauthor{th_ca_Fuji1982} and the $^{28}$Al/Be and $^{206}$Bi/Be measurements of \citeauthor{th_ca_John_Prosser}, all of the reanalyzed photoneutron measurements used homogeneous and isotropic neutron detectors with well modelled and slowly varying sensitivities for low-energy neutrons.  Such detector designs are not suitable for use in a beamline environment.

\citet{BeOxs} and \citet{Utsunomiya2} detected neutrons using hollow cylindrical moderators with thermal neutron detectors positioned in concentric rings within the moderator.  The ratio of the neutron detection rate in each ring allows the measurement of a weighed average neutron energy.  \citet{Arnold_eff} calibrated their detector using several well known neutron production processes, and found a 4\% discrepancy between the simulated and measured ratio of the count rate in the inner versus the outer ring, as shown in their Figure 13.  \citeauthor{Utsunomiya2} claim to have measured a significant number of neutrons from the $^9$Be$(\gamma,n)$ reaction at energies well below those expected from the $^9$Be$(\gamma,n)^8$Be(g.s.) reaction channel.  The calibration of their neutron energy scale depends on simulations that may have a similar discrepancy to that found by \citeauthor{Arnold_eff}.  Their claim likely would have been reported by \citet{BeOxs}, and may also be inconsistent with the neutron energy spectrum measurements of $^{116m}$In/Be and other photoneutron sources using proton recoil spectrometers \cite{photon_rec}.  These inconsistencies demonstrate the importance of the neutron yield measurements from well-modelled detectors used in the present analysis.

\begingroup
\squeezetable
\begin{table}
\centering
\begin{ruledtabular}
\begin{tabular}{c c c c c}
Reference & $E_R$ & $E_{\gamma,\text{ max}}$ & $\sigma_\text{max}$ & FWHM \\
          & (keV) & (keV) & (mb) & (keV) \\
\hline\\[-1.5ex]
Present & $1736.8(18)$ & 1696.8 & 1.43 & 152 \\
\citet{BeOxs} & $1731(2)$ & 1698.3 & 1.64 & 138 \\
\citet{Utsunomiya2} & $1728(1)$ & 1694.3 & 1.35 & 161 \\ 
\citet{th_ca_Burda} & $1748(6)$ & 1705.9 & 1.03 & 177 \\
\end{tabular}
\end{ruledtabular}
\caption{\label{tab:peak} The $^9$Be$(\gamma,n)$ excitation function peak energy, cross section, and width near the $1/2+$ resonance, and the resonance energy for recent experiments.  Note that the position of the excitation function peak and the resonance position are not simply related for this near threshold resonance.  Charged particle scattering experiments including \citet{th_ca_Burda} fail to accurately reproduce the cross section measured by direct experiments.  The peak positions and areas measured by inverse Compton scatter photon beam experiments\cite{BeOxs, Utsunomiya2} matches the present analysis, while the peak widths and heights differ significantly.
}
\end{table}
\endgroup

Table~\ref{tab:peak} shows a comparison of the position, height, and width of the excitation function peak of the $1/2^+$ resonance for radioisotope, inverse Compton, and electron scattering experiments.  The radioisotope experiments find a peak position that agrees well with the measurements of \citeauthor{BeOxs} and \citeauthor{Utsunomiya2}.  The different resonance parameters between these experiments is due to differences in the measured shape of the peak that may be caused by miscalibrated neutron detectors.

\section{Astrophysical Production}

\begin{table*}
\centering
\begin{ruledtabular}
\begin{tabular}{D{.}{.}{4} c D{.}{.}{4} c D{.}{.}{4} c D{.}{.}{3} c}
\multicolumn{1}{c}{T (GK)} & Rate & \multicolumn{1}{c}{T (GK)} & Rate & \multicolumn{1}{c}{T (GK)} & Rate & \multicolumn{1}{c}{T (GK)} & Rate \\
\hline\\[-1.7ex]
0.001 & $1.21(11)\times 10^{-59}$ & 0.016 & $1.04(9)\times 10^{-24}$ & 0.14 & $3.98(12)\times 10^{-8}$ & 1 & $6.22(13)\times 10^{-7}$\\
0.002 & $1.01(9)\times 10^{-47}$ & 0.018 & $8.44(71)\times 10^{-24}$ & 0.15 & $5.87(18)\times 10^{-8}$ & 1.25 & $4.72(9)\times 10^{-7}$\\
0.003 & $6.31(56)\times 10^{-42}$ & 0.02 & $5.20(43)\times 10^{-23}$ & 0.16 & $8.20(24)\times 10^{-8}$ & 1.5 & $3.62(6)\times 10^{-7}$\\
0.004 & $2.89(25)\times 10^{-38}$ & 0.025 & $2.49(19)\times 10^{-21}$ & 0.18 & $1.40(4)\times 10^{-7}$ & 1.75 & $2.83(5)\times 10^{-7}$\\
0.005 & $1.18(10)\times 10^{-35}$ & 0.03 & $3.84(21)\times 10^{-19}$ & 0.2 & $2.10(6)\times 10^{-7}$ & 2 & $2.259(35)\times 10^{-7}$\\
0.006 & $1.16(10)\times 10^{-33}$ & 0.04 & $1.61(8)\times 10^{-15}$ & 0.25 & $4.08(10)\times 10^{-7}$ & 2.5 & $1.527(22)\times 10^{-7}$\\
0.007 & $4.58(40)\times 10^{-32}$ & 0.05 & $2.36(11)\times 10^{-13}$ & 0.3 & $5.92(14)\times 10^{-7}$ & 3 & $1.107(16)\times 10^{-7}$\\
0.008 & $9.58(83)\times 10^{-31}$ & 0.06 & $6.20(28)\times 10^{-12}$ & 0.35 & $7.35(17)\times 10^{-7}$ & 3.5 & $8.47(15)\times 10^{-8}$\\
0.009 & $1.26(11)\times 10^{-29}$ & 0.07 & $6.15(26)\times 10^{-11}$ & 0.4 & $8.32(19)\times 10^{-7}$ & 4 & $6.78(14)\times 10^{-8}$\\
0.01 & $1.16(10)\times 10^{-28}$ & 0.08 & $3.33(13)\times 10^{-10}$ & 0.45 & $8.88(20)\times 10^{-7}$ & 5 & $4.76(15)\times 10^{-8}$\\
0.011 & $8.19(71)\times 10^{-28}$ & 0.09 & $1.21(5)\times 10^{-9}$ & 0.5 & $9.13(20)\times 10^{-7}$ & 6 & $3.62(15)\times 10^{-8}$\\
0.012 & $4.63(40)\times 10^{-27}$ & 0.1 & $3.32(12)\times 10^{-9}$ & 0.6 & $9.00(20)\times 10^{-7}$ & 7 & $2.90(14)\times 10^{-8}$\\
0.013 & $2.19(19)\times 10^{-26}$ & 0.11 & $7.48(26)\times 10^{-9}$ & 0.7 & $8.43(18)\times 10^{-7}$ & 8 & $2.40(13)\times 10^{-8}$\\
0.014 & $8.92(76)\times 10^{-26}$ & 0.12 & $1.45(5)\times 10^{-8}$ & 0.8 & $7.70(16)\times 10^{-7}$ & 9 & $2.05(13)\times 10^{-8}$\\
0.015 & $3.21(27)\times 10^{-25}$ & 0.13 & $2.51(8)\times 10^{-8}$ & 0.9 & $6.94(14)\times 10^{-7}$ & 10 & $1.78(12)\times 10^{-8}$
\end{tabular}
\end{ruledtabular}
\caption{\label{tab:aan} Calculated rate from the two step $^4$He$(\alpha n,\gamma)^9$Be fusion reaction versus temperature.  By convention, rates are in units of (mol/cm$^3$)$^{-2}$s$^{-1}$\cite{th_ca_NACRE}.}
\end{table*}

\begin{figure}
\centering
\includegraphics{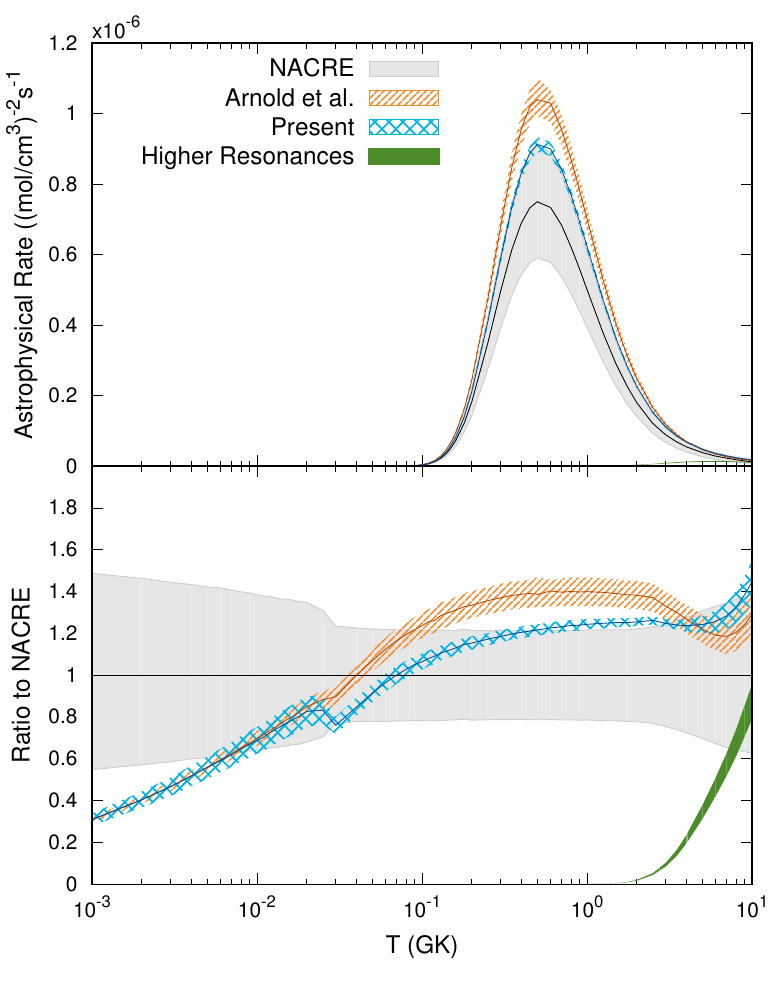}
\caption{\label{fig:aan} Calculated rate from the two step $^4$He$(\alpha n,\gamma)^9$Be fusion reaction compared to previous evaluations \cite{BeOxs, th_ca_NACRE}.  Bands show $\pm1\sigma$ uncertainties.}
\end{figure}

The $^9$Be$(\gamma,n)^8$Be cross section may be used to calculate the dominant two step contribution to the astrophysical production rate of the $\alpha(\alpha n,\gamma)^9$Be reaction.  The reaction proceeds via
\begin{gather*}
\alpha + \alpha \rightarrow ^8\text{Be}  \\
^8\text{Be} + n \rightarrow ^9\text{Be}
\end{gather*}
with both $^8$Be and $^9$Be in their ground state.  A direct three-body reaction is possible and may dominate the reaction rate at low temperatures \cite{Be9_cluster2}, but high-sensitivity searches for this process have not demonstrated its existence \cite{three-body}.  The thermal production rate by the two-step process is a double integral over the energy of two Maxwell-Boltzman distributed collision velocities, the cross sections for the two subprocesses, and the mean lifetime of the $^8$Be nucleus, where the $^8$Be nucleus may be produced off-shell \cite{th_ca_NACRE}.

The $^8$Be$(n,\gamma)^9$Be reaction is related by the reciprocity theorem to the $^{9}$Be$(\gamma,n)^8$Be cross section, as described by \citet{BeOxs}.  The cross section for elastic $\alpha+\alpha$ scattering via the $^8$Be compound state is taken from \cite{aaxs}, with resonance width $\Gamma_{\alpha \alpha} = 5.57 \pm 0.25$~eV and resonant energy $E_r=92.03$~keV.

Table~\ref{tab:aan} shows the adopted $^9$Be astrophysical production rate while Figure~\ref{fig:aan} shows this rate in comparison with the calculations of NACRE \cite{th_ca_NACRE} and \citet{BeOxs}.  At low temperatures, the rate follows that of \citeauthor{BeOxs}.  The low temperature rate differs from that of NACRE by a factor of $\sqrt{E/92.03\text{ keV}}$ in the integrand, where $E$ is the center of mass energy of the $\alpha$ particles in the first reaction step. At astrophysically relevant temperatures of 1~GK to 5~GK, the new rate lies largely between those from NACRE and \citeauthor{BeOxs}.  At high energies, the contribution from higher energy resonances is increased compared to that calculated by \citeauthor{BeOxs} because of increases to the resonance widths (see Table~\ref{tab:upper_res} caption).

\section{Summary}
Radioisotope source yield measurements of the near threshold $1/2^+$ state of the $^9$Be$(\gamma,n)^8$Be reaction cross section have been reanalyzed.  After reanalysis, these measurements are self-consistent and provide precise experimental bounds of the Breit-Wigner parameters, with absolute cross section uncertainties under 2\%.  Twenty percent level inconsistencies between the results of \citet{BeOxs}, \citet{Utsunomiya2}, and this reanalysis are likely due to systematic errors in the low-energy neutron absolute detection efficiency of the two beamline experiments.  The best fit to the cross section indicates that the $1/2^+$ state is a resonance.

\begin{acknowledgments}
This work was derived from my thesis under the guidance of Juan Collar at the University of Chicago. I thank Juan Collar, Charles Arnold and Hiroaki Utsonomiya for discussions about recent measurements of this cross section. This work was completed with the support of the Fermi Research Alliance, LLC under Contract No. De\-AC02\-07CH11359 with the United States Department of Energy. 
\end{acknowledgments}

\bibliography{ref}

\end{document}